# Statistical Analysis on Bangla Newspaper Data to Extract Trending Topic and Visualize Its Change Over Time

**Syed Mehedi Hasan Nirob**[1], **Md. Kazi Nayeem**[1] **and Md. Saiful Islam**[1]
[1] Department of Computer Science and Engineering, Shahjalal University of Science & Technology, Sylhet, Bangladesh**.**
Email: smh.nirob@gmail.com, masum.nayeem@gmail.com, saiful-cse@sust.edu

*Keywords:*

- *News Trend*
- *Newspaper*
- *Frequency*
- *N-gram*
- *Categorization*

**Abstract**: Trending topic of newspapers is an indicator to understand the situation of a country and also a way to evaluate the particular newspaper. This paper represents a model describing few techniques to select trending topics from Bangla Newspaper. Topics that are discussed more frequently than other in Bangla newspaper will be marked and how a very famous topic loses its importance with the change of time and another topic takes its place will be demonstrated. Data from two popular Bangla Newspaper with date and time were collected. Statistical analysis was performed after on these data after preprocessing. Popular and most used keywords were extracted from the stream of Bangla keyword with this analysis. This model can also cluster category wise news trend or a list of news trend in daily or weekly basis with enough data. A pattern can be found on their news trend too. Comparison among past news trend of Bangla newspapers will give a visualization of the situation of Bangladesh. This visualization will be helpful to predict future trending topics of Bangla Newspaper.

## 1. INTRODUCTION

Trending topic analysis is actually to spot a pattern or trend on collected date. Trending topic may vary from time to time and from place to place. It's the people who decide which topic will rule for a certain time or will be most discussed. Current trends for a particular area can be detected with satisfactory precision if we survey on data that represent the current situation of that area. With the expansion of social media like Facebook, Twitter this type of data is becoming more and more available. But newspapers are another option that can help us in this case.

Online content of newspapers is growing rapidly with the change of time and growing audience. Daily newspapers update news on daily basis and people can find their news online. Statistical analysis on these newspaper data can reveal the situation of a particular country.

In Bangladesh, newspapers have always been a popular and important source of information in Bangladesh. Cause in past a large number of people in our country had no internet access. Although at present almost all of the Bangladeshi newspapers have an online version. In this thesis, we are proposing a generalized model to find most discussed topic of Bangla newspaper. We build this model by using data of Prothom Alo[1] and Kaler Kantho[2]. But this model is effective for any Bangla newspaper. Statistical analysis on Bangla language word is the main concern here.

Newspaper trending topic changes with the change of time. Political change, economic and social change, people's behavior etc. A model will be developed that measures the popularity of different subjects in Bangla newspaper and try to find a hidden pattern on these changes.

\* Corresponding author: bangladesh@yourdomain.org

## 2. RELATED WORKS ON TRENDING TOPIC

Trending topic analyses specially find a way to extract most used and meaningful keyword's that is popular among people from the stream of data. News trend analysis for newspaper's written in major languages like English, Spanish, French is not a new concept. But research work was basically done with the data of social media like Twitter, Google, Wikipedia etc. There is a comprehensive study with these three major online and social media stream[3].

L. M. Aiello et al. compared six topic detection methods to sense twitter trends from Twitter datasets related to major events[4]. In 2012 Rong Lu and Qing Yang worked on Twitter trend analysis including reason analysis for news topics[5]. In social media trending topic is not constant and it rises and decay after an amount of time[6]. For a particular time, we can easily take a look at the most popular topic right now on twitter. Real-time trending topic or streaming trends with previous statistics has been done with twitter data[7]. Besides twitter hashtag trends is another effective trend analysis that only work with hash-tagged topic and popularity of them[8]. Real-time classification of Twitter trending topics is another important research work that was done in 2013[9]. Also, these trending topics can be categorized in different subjects like science, sports, politics, technology, music etc. In 2011 Lee K., Palsetia D., Narayanan R., Patwary, M.M.A., Agrawal A. and Choudhary classified Twitter Trending Topics into 18 general categories[10]. They used two approaches for topic classification. These are Bag-of-Words approach for text classification and network-based classification.

But for Bangla language, there is no such work on trending topic analysis. There were some analysis and observation on Bangla news corpus that is not directly related to trending topic analysis[11]. But for working with Bangla language corpus these are important concepts and also there was some statistical analysis on Bangla newspaper data.

## 3. METHODOLOGY

We collected data from two popular Bangladeshi newspaper "Prothom-Alo" and "Kaler Kantho". We separated each word and excluded stop words from them. Then we performed unigram on these data using Chi-squared test[12]. In this process, we placed keywords in top place with higher Chi-squared test value and for same Chi-squared test value, we considered their frequency. We marked them as trending topic candidate for a week. Later for both bigram and trigram, we followed same strategy but this time for two and three consecutive words. Keywords in topmost place became more relevant to the trending topic. Then we picked some keyword those have very high probability of becoming a trending topic and plotted some comparison graphs using those keywords. We manually clustered keywords having similar meaning in our news data to visualize some comparison.

## 4. EXPERIMENT AND PERFORMANCE

### *4.1 Data Collection and Preprocessing*

Data collection has been done using a custom-made web crawler from the online version of two most popular Bangla Newspaper Prothom Alo and Kaler Kantho. It systematically browsed the website of these two newspaper and stored relevant data for this thesis work.

Raw data collected from Prothom Alo are not structured and not appropriate for performing statistical analysis. To make structure data set a custom corpus has been built by the authors containing news from Prothom Alo and Kaler Kantho separated by date.

For each corresponding date, there is a list of word. In every language, there are some common words used more frequently and not very meaningful if compared with keywords. And they can't be trending topic candidate. These stop words were filtered from the corpus.

A list of Bangla stop words[13] was prepared and combined. Final stop word list contains approximately 500 words. After excluding these words from the main corpus, this corpus became more reliable and give better performance. Then a Bangla stemmer was used[14] to find the root word of these words. But, the performance of this stemmer is not that much satisfactory. So, both normal words and root words were considered to perform statistical analysis.

### *4.2 Unigram*

Unigram works with a single word. At first, the frequency of each separated word was counted. This simple unigram or frequency counting gives some interesting result. But most of the word with high frequency are irrelevant and doesn't make sense as a trending topic.

For each keyword, the value of Chi-squared test has been found. The formula is:

$$\frac{(observed - expcted)^2}{expected} \qquad (1)$$

Here in Eq. (1),

$observed = $ Frequency of a word

$expected = $ Avarage frequency of a word

But for different unigram keyword, there is same Chi-square value. In this case, a tie was broken by considering frequency and to plot unigram trends.

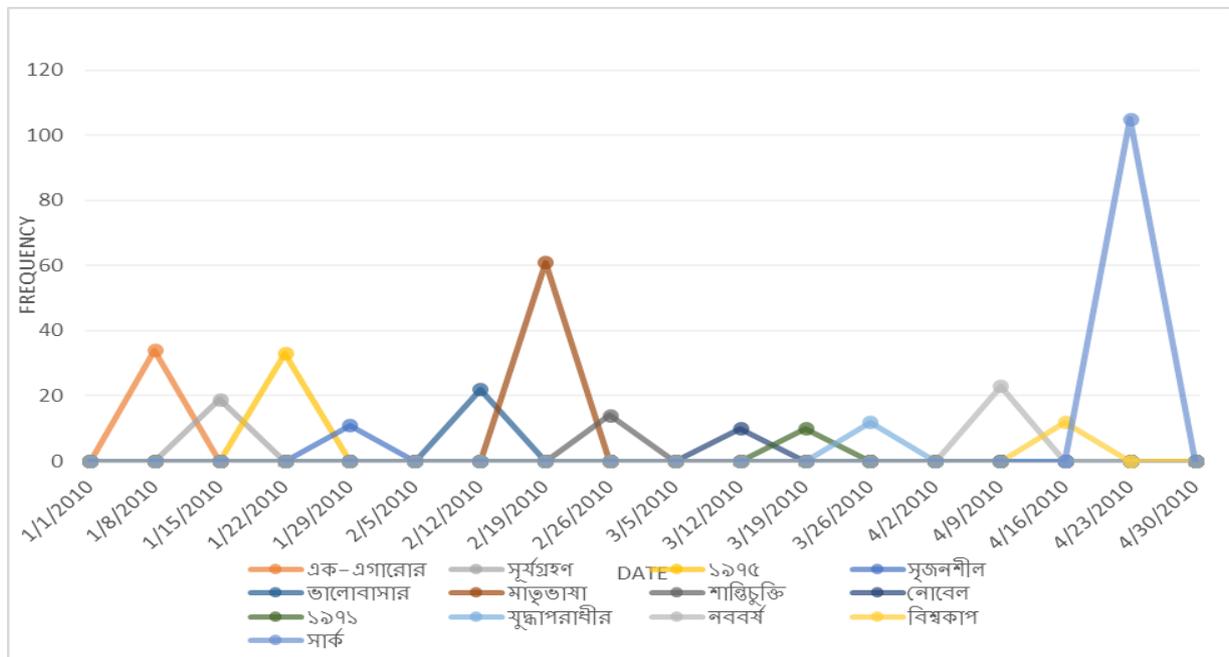

Fig.1   Trending topics of Prothom Alo for every week from January 1, 2010 to April 30, 2010 using unigram

From Fig. 1, Weekly trending topics of Prothom Alo newspaper for 4 months can be shown. Also, a relative popularity of each topic from this graph can be visualized.

### *4.3 Bigram*

After applying unigram on our news data, it can be observed that more than one unigrams had same value in the Chi-squared test. And some trending topic just doesn't make sense and bigrams have been used to ignore them. The frequency of two adjacent keyword or bigrams was counted. Now the Chi-squared test gives distinct value for each bigram and result is also better than unigram.

Table 1 shows Chi-Square test value of top 5 bigrams of a week with frequency. From this table, it can be examined that Chi-squared test value for each bigram is quite different and clearly distinguishable. By using this method top topic of a week can be found. Also, graph (Fig. 2) for some of the top bigrams has been plotted, that show how one topic rise or fall with the change of time.

Table 1: Chi square test value and frequency for top 5 topic from March 25, 2010 to March 31, 2010

| Topic | Chi Square Test Value | Frequency |
|---|---|---|
| মহান স্বাধীনতা | 304.76190476190476 | 29 |
| স্বাধীনতা দিবস | 287.95574387947266 | 34 |
| তদন্ত সংস্থা | 268.32655826558266 | 27 |
| স্বাধীন বাংলা | 240.1 | 27 |
| বিআরটিসির বাস | 192.66666666666669 | 12 |

In Fig. 2, by observing bigram trending topic of Bangla newspaper, it can be showed that they are more pertinent than unigram trending topic. Related data was not clustered. Besides frequency of some unigram is higher but they are not actually trending topic candidate. Some of them can be considered as stop words of Bangla language. For bigrams, this problem diminished dramatically. Accuracy for bigram trending topics is much higher than unigram trending topics.

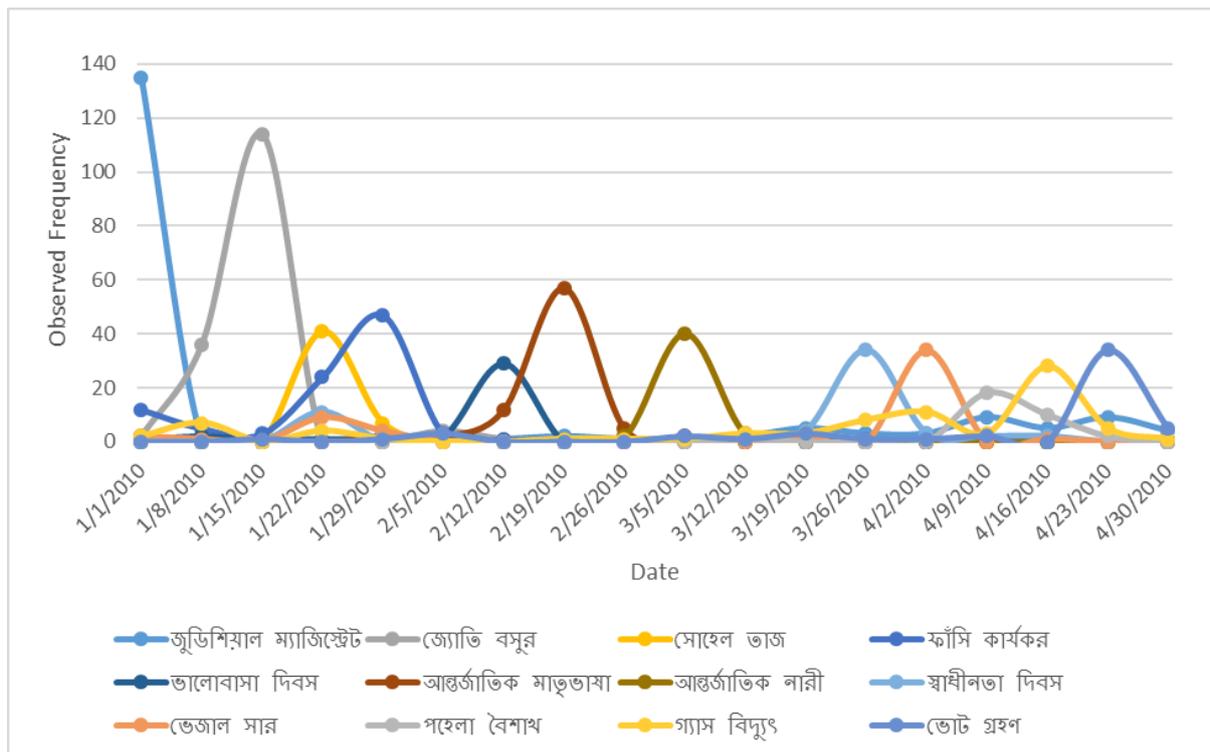

Fig.2 Some Trending topics of Prothom Alo and Kaler Kantho for every week from January 1, 2010 to April 30, 2010 using bigram

*4.4 Trigram*

Frequency measurement of bigrams showed better performance in finding trending topic of Bangla newspaper. Now, the frequency of three adjoining keyword or trigrams has been counted. Similar technique used while finding bigrams trending topic was followed in this case.

From Table 2, that shows top 5 trigrams of a week with their Chi-square test value and frequency, readers can notice that topics with top value and frequency are almost similar to bigram keyword (Table 2).

Table 2: Chi square test value and frequency for top 5 topic from February 19, 2010 to February 25, 2010

| Topic | Chi Square Test Value | Frequency |
|---|---|---|
| শহীদ মিনারে ফুল | 432.7424242424243 | 30 |
| আন্তর্জাতিক মাতৃভাষা দিবস | 348.1 | 32 |
| শহীদ মিনারে পুষ্পার্ঘ্য | 209.1797385620915 | 15 |
| বিমানবন্দরের নাম পরিবর্তন | 180.4738562091503 | 14 |
| মাতৃভাষা দিবস উপলক্ষে | 173.61111111111111 | 15 |

But for some week trigram produces better results than bigram. The peak value of the frequency of same trigrams can also be found in different places (Fig. 3). That's why for non-clustered data trigram is more suitable than unigram or even bigram.

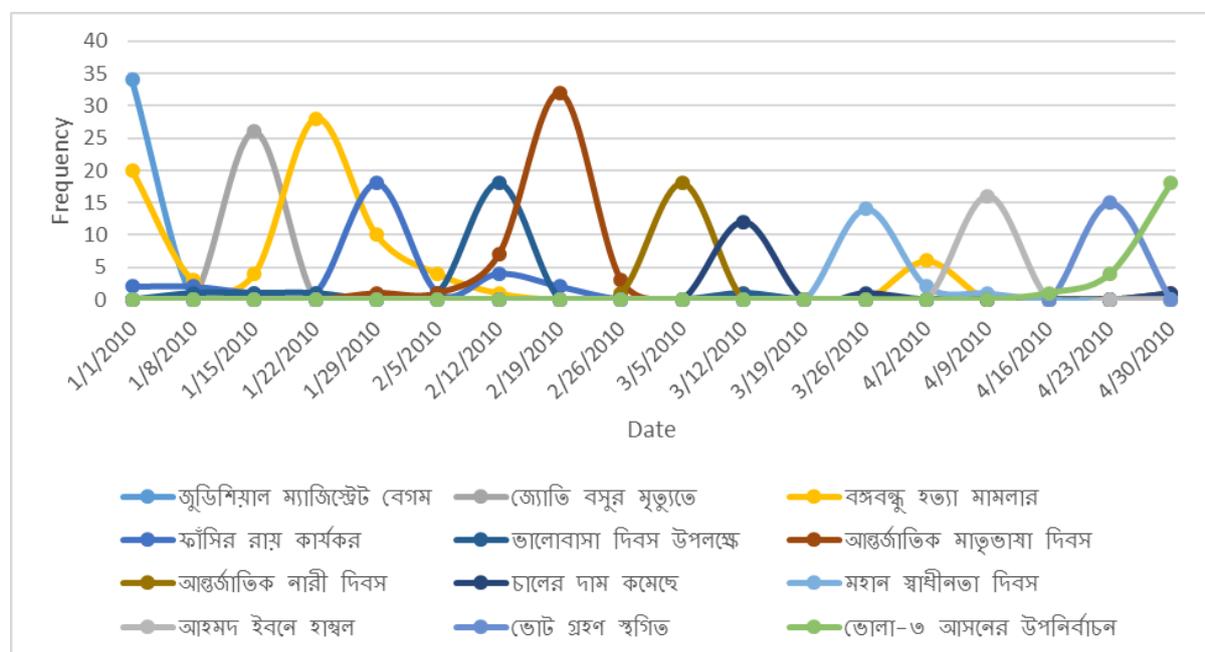

Fig.3 Trending topics of Prothom Alo for every week from January 1, 2010 to April 30, 2010 for trigrams

### 4.5 Analysis and Comparison for Specific Keyword

Newspaper trends can be classified into different categories like Entertainment, Economics, Sports, Education etc. Trending topic for each of these categories may vary. A universal trend can also be selected for a specific time. Now, plotting some graph by providing clustered keywords manually can provide some interesting fact. These keywords can be related to gender, politics, or cultural world. This analysis will reveal the inner philosophy of Bangla newspaper. Do they provide politically biased opinions? Do they represent common people? Answer of these questions will be answered from this specific keyword analysis on Bangla newspaper data.

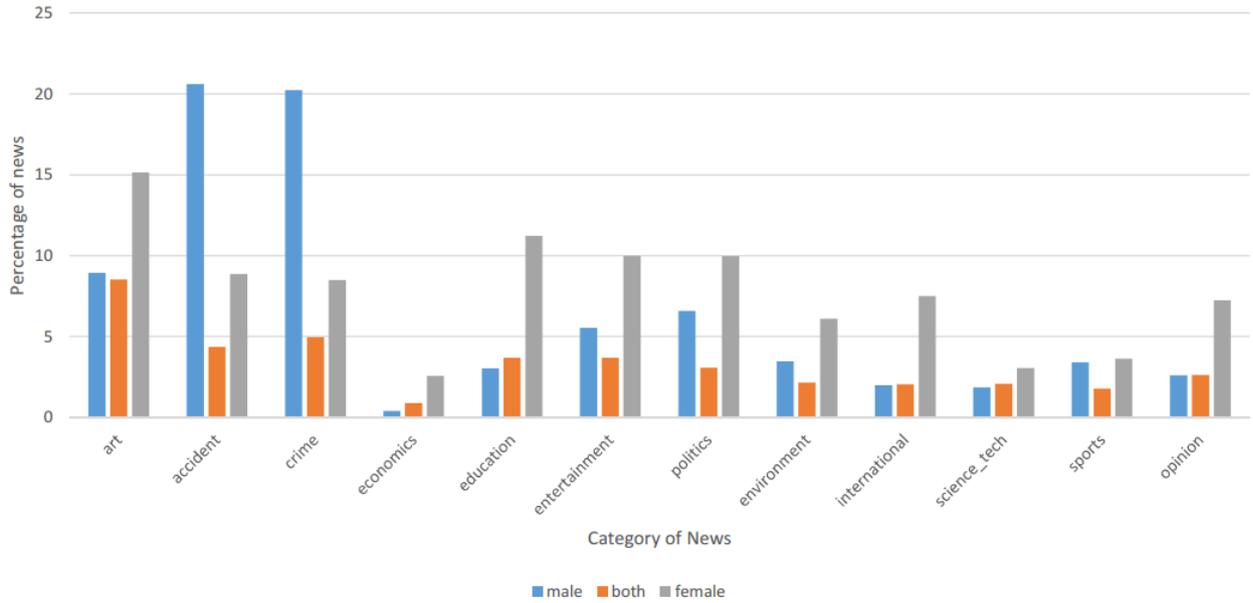

Fig.4   Percentage of news about Men and Women in different category in Prothom Alo Newspaper

The graph in Fig. 4 compares between male and female in different categories. This graph shows that in crime and accident category male is discussed most in "Prothom Alo" newspaper. And in other categories female is discussed more frequently than male. In sports, male and female get almost similar priority. From this graph, it can be said that female get much attention than male in Bangla newspaper.

Table 3:  Closely related 10 words related to women and can be in same cluster

| **Bangla Words** |
| --- |
| নারী |
| নারীশিক্ষা |
| নারীবাদ |
| নারীত্ব |
| মহিলা |
| মেয়ে |
| কন্যা |
| বালিকা |
| স্ত্রী |
| মা |

Table 3 shows closely related words to perform an analysis of male-female comparison. Some of those words are synonyms of each other and some of those words have same contextual meaning. A similar table has been made for male and clustered them.

### *4.6  Performance*

This model to analyze trending topic for Bangla newspaper is the very first model of this kind. So, the result after applying this method was average. Bangla word processing is not an easy task and there is not enough resource to process data and to perform statistical analysis on these data.

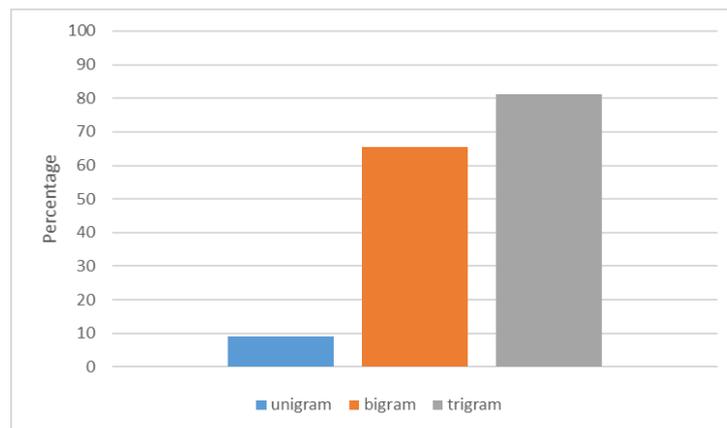

Fig.5:  Percentage of relevant unigrams, bigrams and trigrams trending topics in Bangla newspapers

Statistical analysis has been performed on 18-week news data. Then top 5 trends have been picked for each week. After analyzing the output, it was observed that for unigrams this generalized model doesn't show that much relevant trending topic. Fig. 5 shows the performance of unigrams, bigrams and trigrams trending topic for Bangla newspaper. So, from this analysis, it can be said that for trending topic analysis on Bangla newspaper trigrams show the best performance.

## 5. CONCLUSION

In this paper, a generalized model was designed that can be used to perform trending topic analysis for Bangla newspaper. Trending topics of Bangla newspaper follow some specific pattern. Performing some analysis on this pattern can be helpful to predict future trending topic or future situation of this country. That's why trending topic analysis on Bangla newspaper data is mattering much.

Improvement on this model can make performance much better. Like clustering keywords having similar contextual meaning and then perform statistical analysis on each cluster. Additional improvement on this generalized model will make performance better and much accurate Bangla newspaper trending topic selection will be possible.